\documentclass[twocolumn,showpacs,preprintnumbers,amsmath,amssymb,floatfix]{revtex4}

\usepackage{graphicx}
\usepackage{dcolumn}
\usepackage{bm}

\newcommand{\pom}{\tt I\! P}
\newcommand{\beq}{\begin{equation}}
\newcommand{\eeq}{\end{equation}}

\begin{document}

\title{Diffractive hadroproduction of W$^{\pm}$ and $Z^0$ bosons at high energies}

\author{M. B. Gay Ducati$^a$, M. M. Machado$^a$}
\author{M. V. T. Machado$^b$}

\affiliation{$^a$ High Energy Physics Phenomenology Group, GFPAE,  IF-UFRGS \\
Caixa Postal 15051, CEP 91501-970, Porto Alegre, RS, Brazil}
\affiliation{$^b$ Centro de Ci\^encias Exatas e Tecnol\'ogicas, Universidade Federal do Pampa \\
Campus de Bag\'e, Rua Carlos Barbosa. CEP 96400-970. Bag\'e, RS, Brazil}

\begin{abstract}

Results from a phenomenological analysis of $W$ and $Z$ hard diffractive
hadroproduction at high energies are reported. Using the Regge factorization approach, we consider the recent diffractive parton density functions extracted by the H1 Collaboration at DESY-HERA. In addition, we take into account multiple Pomeron exchange corrections considering a gap survival probability factor. It is find that the ratio of diffractive to non-diffractive boson production is in good agreement with the CDF and D0 data. We make predictions which could be compared to future measurements at the LHC.

\end{abstract}


\maketitle

\section{Introduction}

Recently, the diffractive processes are attracting much attention as a
way of amplifying the physics programme at proton colliders,
including new channels searching for New Physics. The investigation of these reactions at high energies gives
important information on the structure of hadrons and their
interaction mechanisms. Hard diffractive processes, such as the
diffractive production of massive electroweak bosons and dijets, allow the study of the interplay
of small- and large-distance dynamics within QCD. The existence of
a hard scale provides the normalization of the Born term diagram. For boson hadroproduction, single-diffractive dissociation  can occur characterized by the existence of
one large rapidity gap, which can be represented by
 the Pomeron exchange. At high energies, there are important
contributions from unitarization effects and the suppression of the single-Pomeron Born
cross section due to the multi-Pomeron
contributions depends, in general, on the particular
hard process. At the Tevatron energy, $\sqrt s = 1.8$~TeV, the
suppression is in the range 0.05--0.2~\cite{GLM,KMRsoft,BH,KKMR}, whereas for LHC energy, $\sqrt{s}=14$ TeV, the suppression  appears to be of order 0.08--0.1 ~\cite{GLM,KMRsoft,KKMR}. Therefore, the correct treatment of
the multiple scattering effects is crucial for the reliability of the
theoretical predictions of the cross sections for these
diffractive processes.

In the present study, our motivation to perform a new analysis on diffractive boson production is twofold: produce updated theoretical estimations compatible with the current Tevatron data on single diffractive $W$ and $Z$ hadroproduction \cite{CDF,D0} and to perform reliable predictions to the future measurements at the LHC. In order to do so, we use Regge factorization (single Pomeron exchange) and the corresponding corrections for multiple-Pomeron scatterings. Factorization for diffractive hard scattering is equivalent to the hard-scattering aspects of the Ingelman and Schlein model
\cite{IS}, where diffractive scattering is attributed to the
exchange of a Pomeron, i.e. a colorless object with vacuum quantum
numbers. The Pomeron is treated like a real particle,
and one considers that a diffractive electron-proton collision is
due to an electron-Pomeron collision and that a diffractive
proton-proton collision is due to a proton-Pomeron collision.
Therefore, the diffractive hard cross sections are
obtained as a product of a hard-scattering coefficient, a known Pomeron-proton coupling, and parton densities in
the Pomeron. The parton densities in the Pomeron have been sistematicaly extracted from
diffractive DIS measurements. In particular, the quark singlet and gluon content of the Pomeron is obtained from the diffractive struture function $F_{2}^{D(3)}(x_{\pom},\beta,Q^2)$.  Recently, a new analysis of these diffractive parton distributions has been presented \cite{H1diff} by the H1 Collaboration in DESY-HERA. On the other hand, it is well known that the single Pomeron approach produces results that overestimate the experimental values by a large factor \cite{Alvero,mara}. Thus, in the present analysis the corresponding multiple Pomeron exchange corrections will be taken into account.

This paper is organized as follows. In the next section, we present the main formulae to compute the inclusive and diffractive cross sections for $W$ and $Z$ hadroproduction. We show the details concerning the parameterization for the diffractive partons distribution in the Pomeron, extracted recently in DESY-HERA. In addition, we present the theoretical estimations for the gap survival propability factor that will be used in the comparison of our results with experimental measurements from Tevatron and extrapolations to the LHC energy.  In the last section we present our numerical results and perform predictions to future measurements in CERN LHC experiment. The compatibility with data is analysed, possible additional corrections are investigated and the comparison with other approaches are  considered.

%
%
\section{Diffractive Hadroproduction of Massive Gauge Bosons}

Let us start by introducing the main expressions to compute the inclusive and diffractive cross sections.
For the {\it hard} diffractive
processes we will consider the Ingelman-Schlein (IS) picture \cite{IS}, where
the Pomeron structure (quark and gluon content) is probed. The starting point is the generic cross section for a process in which partons of two hadrons, $A$ and $B$, interact to produce a massive electroweak boson, $
 A + B \rightarrow (W^{\pm}/Z^0) + X$, 
\begin{eqnarray}
\nonumber
\frac{d \sigma}{dx_a\,dx_b} = \sum_{a,b} f_{a/A}(x_a,
\mu^2)\, f_{b/B}(x_b, \mu^2)\, \frac{d\hat{\sigma}(ab\rightarrow [W/Z]\,X)}{d\hat{t}}\,,
\label{gen}
\end{eqnarray}
where $x_i f_{i/h}(x_i, \mu^2)$ is the parton distribution function of a parton of flavour $i=a,b$ in the hadron $h=A,B$.  The quantity $d\hat{\sigma}/d\hat{t}$ gives the elementary hard cross section of the corresponding subprocess and $\mu^2=M_{W/Z}^2$ is the hard scale in which the pdf's are evolved in the QCD evolution. Equation above express the usual leading-order QCD procedure to obtain the non-diffractive cross section. Next-to-leading-order
contributions are not essential for the present purposes.

In order to obtain the corresponding expression for
diffractive processes, one assumes that one of the hadrons, say
hadron $A$, emits a Pomeron whose partons interact with partons of the hadron $B$.
Thus the parton distribution  $x_a f_{a/A}(x_a, \mu^2)$ in
Eq.~(\ref{gen}) is replaced by the convolution between a 
distribution of partons in the Pomeron, $\beta f_{a/{\pom}}(\beta,
\mu^2)$, and the ``emission rate" of Pomerons by the hadron, $f_{{\pom}/h}(x_{{\pom}},t)$. The last quantity, $f_{{\pom}/h}(x_{{\pom}},t)$, is the Pomeron flux factor and its explicit formulation is described in
terms of Regge theory. Therefore, we can rewrite the parton distribution as
\begin{eqnarray}
\nonumber
\label{convol}
x_a f_{a/A}(x_a, \,\mu^2) & =& \int dx_{{\pom}} \int d\beta \int dt\,
f_{{\pom}/A}(x_{{\pom}},\,t) \\
&\times & \beta \, f_{a/{\pom}}(\beta, \,\mu^2)\,
\delta \left(\beta-\frac{x_a}{x_{{\tt I\! P}}}\right),
\end{eqnarray}
and, now defining $\bar{f} (x_{{\pom}}) \equiv \int_{-\infty}^0 dt\
f_{{\pom/A}}(x_{{\pom}},t)$, one obtains
\begin{eqnarray}
\label{convoP}
x_a f_{a/A}(x_a, \,\mu^2)\ =\ \int dx_{{\pom}} \
\bar{f}(x_{{\pom}})\, {\frac{x_a}{x_{{\pom}}}}\, f_{a/{\pom}}
({\frac{x_a}{x_{{\pom}}}}, \mu^2).
\end{eqnarray}

Concerning the $W^{\pm}$ diffractive production, one considers the reaction
$p + {\bar p}(p) \rightarrow p + \ W (\rightarrow e\ \nu ) + \ X$, assuming that a Pomeron emitted by a proton in
the positive $z$ direction interacts with a $\bar p$ (or a $p$) producing $W^{\pm}$
that subsequently decays into $e^{\pm}\ \nu$. The detection of this reaction is
triggered by the lepton ($e^+$ or
$e^-$) that appears boosted towards negative $\eta$ (rapidity) in coincidence
with a rapidity gap in the right hemisphere.

By using the same concept of the convoluted structure function, the
diffractive (single diffraction, SD) cross section for the inclusive lepton production for
this process becomes
\begin{widetext}
\begin{eqnarray}
\frac{d\sigma^{\mathrm{SD}}_{\mathrm{lepton}}}{d\eta_e}= \sum_{a,b}
\int \frac{dx_{\pom}}{x_{\pom}}\, \bar{f}(x_{\pom})
\int dE_T \ f_{a/{\pom}}(x_a, \,\mu^2)\,f_{b/\bar{p}(p)}(x_b, \,\mu^2)\
\left[\frac{ V_{ab}^2\ G_F^2}{6\ s\ \Gamma_W}\right]\ \frac{\hat{t}^2}
{\sqrt{A^2-1}}
\label{dsw}
\end{eqnarray}
\end{widetext}
where
\begin{equation}
x_a = \frac{M_W\ e^{\eta_e}}{(\sqrt{s}\ x_{{\tt I\! P}})}\ \left[A \pm
\sqrt{(A^2-1)}\right],
\label{xaw}
\end{equation}
\begin{equation}
x_b = \frac{M_W\ e^{-\eta_e}}{\sqrt{s}}\ \left[A \mp \sqrt{(A^2-1)}\right],
\label{xbw}
\end{equation}
and
\begin{equation}
\hat{t}=-E_T\ M_W\ \left[A+\sqrt{(A^2-1)}\right]
\label{tw}
\end{equation}
with $A={M_W}/{2 E_T}$, $E_T$ being the lepton transverse energy, $G_F$ is the Fermi constant and the hard scale $\mu^2=M_W^2$. The quantity  $V_{ab}$ is equal
 to the Cabibbo-Kobayashi-Maskawa matrix element if $e_a+e_{b}=\pm 1$ and zero otherwise, where $a,b$ denote quark flavors and $e_q$ the fractional charge of quark $q$. The upper signs in Eqs.~(\ref{xaw}) and (\ref{xbw})
refer to $W^+$ production (that is, $e^+$ detection). The corresponding
cross section for $W^-$ is obtained by using the lower signs and ${\hat t}
\leftrightarrow {\hat u}$.

In a similar way, the cross section for the diffractive hadroproduction of neutral weak vector boson $Z$
is given by
\begin{widetext}
\begin{eqnarray}
  \sigma ^{\mathrm{SD}}_{Z}\,(\sqrt{s}) = \sum _{a,b} \int \frac{dx_{\pom}}{x_{\pom}}
    \int \frac{dx_b}{x_b} \int \frac{dx_a}{x_a}
    \, \bar{f}(x_{\pom})
    \, f_{a/{\pom}}(x_{a},\,\mu^2)
    \, f_{b/\bar{p}(p)}(x_{b}, \,\mu^2)\left[\frac{2\pi C_{ab}^ZG_FM_Z^2}{3\,\sqrt{2}\, s}\right]\,\frac{d\hat{\sigma}(ab\rightarrow ZX)}
{d\hat{t}}\,,
\label{dsz}
\end{eqnarray}
\end{widetext}
where $C^Z_{q{\bar q}}=1/2-2|e_q|\sin^{2}\theta _{W}
    +4|e_q|^{2}\sin^{4}\theta _{W}$, with $\theta _{W}$ being the Weinberg or weak-mixing angle. The definitions for $x_{a,b}$ are similar as for the $W$ case and now $\mu^2=M_Z^2$. The values of the electroweak parameters
that appear in the various formulae were taken from the particle data
handbook \cite{PDG}, and we use only four flavors ($u$, $d$, $s$, $c$)
in the weak mixing matrix, with the Cabibbo angle $\theta _{C} = 0.2269$.

\subsection{The Pomeron Flux Factor}

An important element in the calculation of hard diffractive cross sections is the Pomeron flux factor, introduced in
Eq.~(\ref{convol}). We take the experimental analysis of the diffractive structure function \cite{H1diff}, where the $x_{\pom}$ dependence is parameterised using a flux factor
motivated by Regge theory \cite{Collins},
\begin{eqnarray}
f_{\pom/p}(x_{\pom}, t) = A_{\pom} \cdot
\frac{e^{B_{\pom} t}}{x_{\pom}^{2\alpha_{\pom} (t)-1}} \ ,
\label{eq:fluxfac}
\end{eqnarray}
where  the Pomeron trajectory is assumed to be linear,
$\alpha_{\pom} (t)= \alpha_{\pom} (0) + \alpha_{\pom}^\prime t$, and the parameters
$B_{\pom}$ and $\alpha_{\pom}^\prime$ and their uncertainties are obtained from
fits to H1 FPS data \cite{H1FPS}. The normalisation parameter $A_{\pom}$ is chosen such that
$x_{\pom} \cdot \int_{t_{\rm cut}}^{t_{\rm min}} f_{\pom/p} \ {\rm d} t
= 1$ at $x_{\pom} = 0.003$, where
$|t_{\rm min}| \simeq m_p^2 \, x_{\pom}^2 \, / \, (1 - x_{\pom})$ is the minimum
kinematically accessible value of $|t|$, $m_p$ is the proton mass and
$|t_{\rm cut}|= 1.0 \rm\ GeV^{2}$ is the limit of the measurement.

The flux factor above corresponds to the standard Pomeron flux from Regge phenomenology, based on the Donnachie-Landshoff model \cite{DLflux}. On the other hand, there is an alternative Pomeron flux, proposed first by Goulianos \cite{Goulianos}, which considers it as a probability density. Thus, the integral over the diffractive phase space could not exceed the unit and the standard flux should be normalized. For instance, see Refs. \cite{mara} for previous phenomenology using the normalized flux in boson hadroproduction.

\begin{figure}[t]
\includegraphics[scale=0.47]{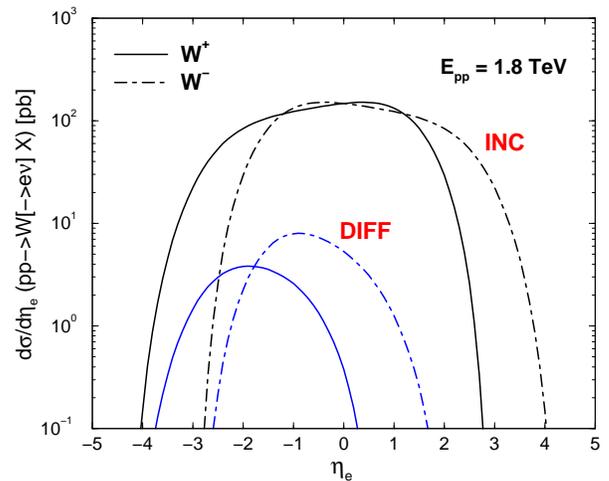}
\caption{(Color online) The rapidity distribution of electron and positron generated in inclusive and diffractive $W$ hadroproduction at $\sqrt{s}=1.8$ TeV (see text).}
\label{fig:1}
\end{figure}

\subsection{The Pomeron Structure Function}

In the estimates for the diffractive cross sections, we will consider the diffractive pdf's recently obtained by the H1 Collaboration at DESY-HERA \cite{H1diff}. The Pomeron structure function has been modeled in terms of a
light flavour singlet distribution $\Sigma(z)$, consisting of $u$, $d$ and $s$
quarks and anti-quarks
with $u=d=s=\bar{u}=\bar{d}=\bar{s}$,
and a gluon distribution $g(z)$.  Here, $z$ is the longitudinal momentum
fraction of the parton entering the hard sub-process
with respect to the diffractive
exchange, such that $z=\beta$
for the lowest order quark-parton model process,
whereas $0<\beta<z$ for higher order processes.
The quark singlet and gluon distributions are parameterised
at $Q_0^2$ with the general form,
\begin{equation}
z f_i (z,\,Q_0^2) = A_i \, z^{B_i} \, (1 - z)^{C_i} \,\exp\left[{- \frac{0.01}{(1-z)}}\right]\,
\label{param:general}
\end{equation}
where the last exponential factor ensures that the diffractive pdf's vanish
at $z = 1$. The charm and beauty quarks are treated as massive, appearing via boson gluon fusion-type processes up to order $\alpha_s^2$. To determine experimentally the diffractive pdf's, the following cuts have been considered: $\beta < 0.8$, $M_X > 2$ GeV and $Q^2 < 8.5 $ GeV$^2$, mostly in order to avoid regions influenced
by higher twist contributions or large theoretical uncertainties \cite{H1diff}.

For the quark singlet distribution,
the data require the inclusion of all three parameters
$A_q$, $B_q$ and $C_q$ in equation~\ref{param:general}.
By comparison, the gluon density is weakly
constrained by the data, which are found to be
insensitive to the $B_g$ parameter. The gluon density is thus
parameterised at $Q_0^2$ using only the $A_g$ and $C_g$ parameters.
With this parameterisation, one has the value  $Q_0^2 = 1.75 \ {\rm GeV^2}$ and it is referred to as the `H1 2006 DPDF Fit A'. It is verified that the fit procedure is not sensitive to the gluon pdf and a new adjust was done with $C_g=0$. Thus, the gluon density is then a simple constant at the starting scale for evolution, which was chosen to be
$Q_0^2 = 2.5 \ {\rm GeV^2}$ and it is referred to as the
`H1 2006 DPDF Fit B'. The quark singlet distribution is well constrained,
with an uncertainty of typically $5 -10\%$ and good agreement between
the results of both fits \cite{H1diff}.

\subsection{The Gap Survival Factor}
In the following analysis we will consider the suppression of the hard diffractive cross section by multi-Pomeron scattering effects.
 This is taken into account through a gap survival probability factor. There has been large interest in the probability of rapidity gaps in high energy interactions to survive as they may be populated by secondary particles generated by rescattering processes. This effect can be described in terms of screening or absorptive corrections, which can be estimated using the quantity \cite{Bj}:
 \begin{eqnarray}
<\!|S|^2\!>=\frac{\int|{\cal{A}}\,(s,b)|^2\,e^{-\Omega (s,b)}\,d^2b}{\int|{\cal{A}}\,(s,b)|^2\,d^2b}\,,
 \end{eqnarray}
where $\cal{A}$ is the amplitude, in the impact parameter space, of the particular process of interest at center-of-mass energy $\sqrt{s}$. The quantity $\Omega$ is the opacity (or optical density) of the interaction of the incoming hadrons. This suppression factor of a hard process accompanied by a rapidity gap depends not only on the probability of the initial state survive, but is sensitive to the spatial distribution of partons inside the incoming hadrons, and thus on the dynamics of the whole diffractive part of the scattering matrix.

For our purpose, we consider two theoretical estimates for the suppression factor. The first one is the work of Ref. \cite{KKMR} (labeled KMR), which considers a two-channel eikonal model that embodies pion-loop insertions in the pomeron trajectory, diffractive dissociation and rescattering effects. The survival probability is computed for single, central and double diffractive processes at several energies, assuming that the spatial distribution in impact parameter space is driven by the slope $B$ of the pomeron-proton vertex. We will consider the results for single diffractive processes with $2B=5.5$ GeV$^{-2}$ (slope of the electromagnetic proton form factor) and without $N^*$ excitation, which is relevant to a forward proton spectrometer (FPS) measurement. Thus, we have $<\!|S|^2\!>_{\mathrm{KMR}}=0.15$ for $\sqrt{s}=1.8$ TeV (Tevatron) and $<\!|S|^2\!>_{\mathrm{KMR}}=0.09$ for $\sqrt{s}=14$ TeV (LHC).

The second theoretical estimate for the gap factor is from Ref. \cite{GLMrev} (labeled GLM), which considers a single channel eikonal approach. We take the case where the soft input is obtained directly from the measured values of $\sigma_{tot}$, $\sigma_{el}$ and hard radius $R_{H}$. Then, one has $<\!|S|^2\!>_{\mathrm{GLM}}=0.126$ for $\sqrt{s}=1.8$ TeV (Tevatron) and $<\!|S|^2\!>_{\mathrm{GLM}}=0.081$ for $\sqrt{s}=14$ TeV (LHC). We quote Ref. \cite{GLMrev} for a detailed comparison between the two approaches and further discussions on model dependence of inputs and consideration of multi-channel calculations.
It should be stressed that our particular choice by KMR and GLM (single channel)  models is in order to indicate the uncertainty (model dependence) of the soft interaction effects. It is worth to mention that some implementations of GLM model include the results of a two or three channel calculation for  $<\!|S|^2\!>$, which are considerably smaller than the one channel result \cite{GLMrev}.

%
%

\section{Results and Discussion}

\begin{table}
\caption{\label{tab:table1} Data versus model predictions for diffractive $W^{\pm}$ hadroproduction (cuts $E_{T_{\mathrm{min}}}=20$ GeV and $x_{\pom}<0.1$). }
\begin{ruledtabular}
\begin{tabular}{lccr}
$\sqrt{s}$  & Rapidity  & Data (\%) & Estimate (\%)\\
\hline
1.8 TeV &    $|\eta_e|<1.1$  & $1.15\pm 0.55$ \cite{CDF}  & $0.715\pm 0.045$\\
1.8 TeV &    $|\eta_e|<1.1$  & $1.08\pm 0.25$ \cite{D0} & $0.715\pm 0.045$\\
1.8 TeV &    $1.5<|\eta_e|<2.5$  & $0.64\pm 0.24$  \cite{D0} & $1.7\pm 0.875$\\
1.8 TeV &     Total $W\rightarrow e\nu $ & $0.89\pm 0.25$ \cite{D0} & $0.735\pm 0.055$ \\
14 TeV &    $|\eta_e|<1$  & ---  & $31.1\pm 1.6$\\
\end{tabular}
\end{ruledtabular}
\end{table}

In the following, we present our predictions for hard diffractive production of
W's and Z's based on the previous discussion. These predictions are compared
with experimental data from Refs.~\cite{CDF,D0} in Tables I and II. In addition, estimations for the LHC are presented. In the numerical calculations, we have used the new H1 parameterizations for the diffractive pdf's \cite{H1diff}. The `H1 2006 DPDF Fit A' was considered and one verifies that the results are not quite sensitive  to a replacement by `H1 2006 DPDF Fit B'. For the usual pdf's in the proton (anti-proton) we have considered the updated MRST2004F4 parameterization \cite{mrst2004f4}, which is a four-fixed-flavour version of the standard MRST2004 parton distributions. As the larger uncertainty comes from the gap survival factor, the error in the predictions correspond to the theoretical band for $<\!|S|^2\!>$. In the theoretical expressions of previous section  only the interaction of pomerons (emitted by protons) with antiprotons (protons in LHC case) are computed, that means events with rapidity gaps on the side from which antiprotons come from. The experimental rate is for both sides, that is events with a rapidity gap on the proton or antiproton side. Therefore, we have multiplied the theoretical prediction by a factor 2 in order to compare it with data.

Let us start by the diffractive $W$ production. In order to illustrate our investigation, in Fig. \ref{fig:1} we present the rapidity distribution of the electron (dot-dashed lines) and positron (solid lines) generated in both inclusive and diffractive $W^{\pm}$ hadroproduction in Tevatron for $\sqrt{s}=1.8$ TeV. The diffractive cross sections are not corrected by gap survival factor and they are given by Eq. (\ref{dsw}). In this case, the diffractive production rate is approximately 7 \% (using the cut $|\eta|<1$) being very large compared to the Tevatron data. When considering the gap survival probability correction, the values are in better agreement with data. When considering central $W$ boson fraction, $-1.1<\eta_e<1.1$ (cuts of CDF and D0 \cite{CDF,D0}), we obtain a diffractive rate of 0.67 \% using the KMR estimate for $<\!|S|^2\!>$, whereas it reaches 0.76\% for the GLM estimate. The average rate considering the theoretical band for the gap factor is then $R_W= 0.715\pm 0.045$ \%. This result is consistent with the experimental central values $R_W^{\mathrm{CDF}}=1.15$ \% and $R_W^{\mathrm{D0}}=1.08$ \%. The agreement would be better whether the sub-leading reggeon contribution is added, which was not considered in present calculation. In Ref. \cite{CFL}, it was shown that its introduction considerably enhances the diffractive ratio in the Tevatron regime. Considering the forward $W$ fraction, $1.5<|\eta_e|<2.5$ (D0 cut), one obtains $R_W=0.83$ \% for KMR and $R_W=2.58$ \% for GLM, with an averaged value of $R_W= 1.7\pm 0.875$ \%. In this case, our estimate is larger than the central experimental value $R_W^{\mathrm{D0}}=0.64$ \%. For the total $W\rightarrow e\nu$ we have $R_W=0.68$ \% for KMR and $R_W=0.79$ \% for GLM  and the mean value $R_W= 0.735 \pm 0.055$ \%, which is in agreement with data and consistent with a large forward contribution. Finally, we estimate the diffractive ratio for the LHC energy, $\sqrt{s}=14$ TeV. In this case we extrapolate the pdf's in proton and diffractive pdf's in Pomeron to that kimenatical region. This procedure introduces somewhat additional uncertainties in the theoretical predictions. We take the conservative cuts $|\eta_e|<1$, $E_{T_{\mathrm{min}}}=20$ GeV for the detected lepton and $x_{\pom}<0.1$. We find $R_W=32.7$ \% for KMR gap survival probability factor and $R_W=29.5$ \% for GLM, with a mean value of $R_W^{\mathrm{LHC}}=31.1\pm 1.6 $ \%. This means that the diffractive contribution reaches one third, or even more, of the inclusive hadroproduction even when multi-Pomeron scattering corrections are taken into account. The reason of this enhancement is the increasingly large diffractive cross section. The results presented above are summarized in Table I. The experimental errors have been summed into quadrature.

\begin{table}
\caption{\label{tab:table2} Data versus model predictions for diffractive $Z^0$ hadroproduction (cuts $E_{T_{\mathrm{min}}}=16$ GeV and $x_{\pom}<0.1$).}
\begin{ruledtabular}
\begin{tabular}{lccr}
$\sqrt{s}$  & Rapidity  & Data (\%) & Estimate (\%)\\
\hline
1.8 TeV &   Total $Z\rightarrow e^+e^-$  & $1.44\pm 0.80$ \cite{D0}  & $0.71\pm 0.05$\\
14 TeV &    Total $Z\rightarrow e^+e^-$  & ---  & $30.26\pm 1.41$\\
\end{tabular}
\end{ruledtabular}
\end{table}

Now, we present the investigations for the diffractive $Z$ hadroproduction.
When the gap survival factor is not considered the diffractive cross section is given by Eq. (\ref{dsz}), producing a diffractive rate of 6.2 \%. This value is once again higher than the Tevatron data by a factor five. When considering the gap survival correction, we verify an agreement with experiment. For the total $Z\rightarrow e^+e^-$ we obtain a diffractive rate of 0.66 \% using the KMR estimate for $<\!|S|^2\!>$, whereas it reaches 0.76 \% for the GLM estimate. The average value gives $R_Z=0.71\pm 0.05$, which is consistent with the experimental result $R_Z^{\mathrm{D0}}=1.44+0.61-0.52$. A rough extrapolation to LHC energy gives $R_Z=31.67$ with KMR gap factor and $R_Z=28.85$ for GLM, with a mean value $R_Z^{\mathrm{LHC}}=30.26\pm 1.41$. We again consider the conservative cuts $E_{T_{\mathrm{min}}}=16$ GeV and $x_{\pom}<0.1$. This estimate follows similar trend as for the $W$ case. The results presented above are summarized in Table II. The experimental errors have been summed into quadrature.

Our results can be compared with previous calculations in diffractive boson hadroproduction. For instance, in Refs. \cite{mara} one uses IS approach with a normalized Pomeron flux \cite{Goulianos} and the corresponding diffractive pdf's. The data description for the $W$ case is reasonable. However, the calculations are only compared to the CDF \cite{CDF} data and they are somewhat larger than ours. In Ref. \cite{CFL} a hard Pomeron flux is considered, i.e. $\alpha_{\pom}(0)\simeq 1.4$, and multiple scatterings are taken into account by a Monte Carlo calculation. In addition, for Tevatron energies the reggeon contribution is added. The results are compared only to CDF data \cite{CDF} for the $W$ production and the description is consistent with experiment. It is interesting the fact that a hard Pomeron flux could mimic the multi-Pomeron suppression  or the effect of normalizing the standard Pomeron flux. Finally, we need call attention to the uncertainty in the determination of the gap survival probability. The estimates considered here (KMR and GLM) are compatible with each other for the case of single diffractive processes. However, recent calculations using one channel eikonal model give larger values for  $<\!|S|^2\!>$ \cite{BH,Luna}. For instance, in Ref. \cite{Luna} an eikonal QCD model with a dynamical gluon mass (DGM) was considered. Using a gluon mass $m_g=400$ MeV, one obtains $<\!|S|^2\!>_{\mathrm{DGM}}(\mathrm{Tevatron})=27.6\pm7.8$~\%  and
$<\!|S|^2\!>_{\mathrm{DGM}}(\mathrm{LHC})=18.2\pm7.0$~\%. These values give $R_W (\sqrt{s}=1.8 \,\mathrm{TeV})\simeq 1.23$~\% and $R_Z (\sqrt{s}=1.8 \,\mathrm{TeV})\simeq 1.21$ \%. This illustrates the size of uncertainty when considering different estimates for the gap probability.

In summary, we have shown that it is possible to obtain a reasonable
overall description of hard diffractive hadroproduction of massive gauge bosons by the model based on Regge factorization supplemented by gap survival factor. For the Pomeron model, we take the recent  H1 diffractive parton density functions extracted from their measurement of $F_2^{D(3)}$. The results are directly dependent on the quark singlet distribution in the Pomeron. We did not observe large discrepancy in using the different fit procedure for diffractive pdf's (fit A and B). We estimate the multiple interaction corrections taking the theoretical prediction of distinct multi-channel models, where the gap factor decreases on energy. That is,  $<\!|S|^2\!>\simeq 15-17.5$ \% for Tevatron energies going down to $<\!|S|^2\!>\simeq 8.1-9$ \% at LHC energy. We find that the ratio of diffractive to non-diffractive boson production is in good agreement with the CDF and D0 data when considering these corrections. The overall diffractive ratio for $\sqrt{s}=1.8$ TeV (Tevatron) is of order 1 \%. In addition, we make predictions which could be compared to future measurements at LHC. The estimates give large rates of diffractive events, reaching values higher than 30~\% of the inclusive cross section.
 

\section*{Acknowledgments}

This work was supported by CNPq, Brazil. M.M. Machado thanks E.G.S. Luna for useful discussions about gap survival probability calculations. The authors thank M. Arneodo and M. Albrow for useful comments. 


%
%


\end{document}